\documentclass[aps,prl,floatfix,superscriptaddress,unsortedaddress,twocolumn,showpacs]{revtex4}
\usepackage{amsmath,amssymb}
\usepackage{soul}
\usepackage{pifont}

\providecommand{\U}[1]{\protect\rule{.1in}{.1in}}
\newcommand{\sign}{\mathrm{sign}}
\newcommand{\dg}{^{\dagger}}
\newcommand{\ua}{_{\uparrow}}
\newcommand{\da}{_{\downarrow}}

\renewcommand{\epsilon}{{\varepsilon}}

\begin{document}
\title{Fully spin-dependent boundary condition for isotropic quasiclassical Green's functions} 
\author{P. Machon}
\affiliation{Department of Physics, University of Konstanz, D-78457 Konstanz, Germany}
\author{W. Belzig}
\affiliation{Department of Physics, University of Konstanz, D-78457 Konstanz, Germany}
\date{\today}
\begin{abstract}
Transport in superconducting heterostructures is very successfully described with quasiclassical Green's functions augmented by microscopically derived boundary conditions. However, so far the spin-dependence is in the diffusive approach included only for limiting cases.
Here, we derive the fully spin-dependent boundary condition completing the Usadel equation and the circuit theory.
Both, material specific spin-degrees of freedom and spin-dependent interface effects, i.e. spin-mixing and polarization of the transmission coefficients are treated exactly.
This opens the road to accurately describe a completely new class of mesoscopic circuits including materials with strong intrinsic magnetic structure. We also discuss several experimentally relevant cases like the tunnel limit, a ferromagnetic insulator with arbitrarily strong magnetization and the limit of small spin-mixing.
\end{abstract}
\maketitle

A central question in quantum transport is to find boundary conditions
(BC) defining the current across contacts of different materials, like
e.g. superconducting heterostructures possibly including ferromagnetic
materials. A straight forward approach to this is to match the
wavefunctions at the interface (see e.g. \cite{btk}). A
generalization of this approach containing many transport channels and
all kinds of different barriers can be formulated with a scattering
matrix \cite{andreev}. However, due to the huge variety of effects in
quantum transport it is in many cases more convenient to work with
Green's functions (see e.g. \cite{rammer:86}) instead of a wave
functions. These Green's functions are described by the Gor'kov
equations \cite{gorkov:58}, which are not practically
solvable in the presence of interfaces and disorder. In the
quasiclassical limit (i.e. the Fermi energy is much larger than all
relevant energy scales) \cite{eilenberger:68} it is however possible
to split the total Green's into a ballistic part that is continuous at
the interface and can thus be described within a scattering approach
and a discontinuity \cite{zaitsev:84}. In this way, BCs for quasiclassical Green's functions can be formulated. 
A generalization to this approach for
magnetically active interfaces is given in \cite{rainer:88}, leading
to a pair breaking effect of a magnetic insulator due to the
spin-mixing \cite{tokuyasu:88}.  Junctions including strongly
diffusive superconductors are treated in\cite{kuprianov:88}. A more
detailed overview can be found in \cite{buzdin:05,bergeret:05}, where
the need for a BC the covers superconductor-strong ferromagnet
interfaces was pointed out.

As already implied, for many physical systems like superconducting
heterostructures mainly two limits are considered. The clean
(ballistic) limit that assumes the transport to be described by
ballistic trajectories and the dirty (diffusive) limit where the
mean-free path is the smallest length scale in the system and the
transport inside the leads is dominated by the impurity self-energy
from Born's approximation.  A BC for stationary problems that covers
spin-dependent effects based on a scattering approach within the clean
limit is given in \cite{eschrig:04}. The aim of this paper is to
provide a fully derived, general BC within the framework of
quasiclassical Green's functions in the case
of strongly diffusive materials. The dirty limit BC for the
spin-independent case is already known since long
\cite{nazarov:99}. In this theory (called quantum circuit theory) a
matrix-current in Nambu-Keldysh space \cite{nambu:55,keldysh:65} is
defined for which generalized Kirchhoff's rules apply as a direct
consequence of the discretization of the Usadel equation
\cite{usadel:70}. Observable currents (like charge, energy and spin
currents) are then defined from the energy integral over subtraces of
the matrix-current. An extension of these BC's that consider
spin-dependent effects to first order in the tunneling is provided in
\cite{huertas:02,cottet:09,machon:13}. Another extension that doesn't consider
spin mixing but an interface polarization is found in
\cite{bergeret:12}. 
However, a BC that treats all the spin-dependent
effects exactly is still missing. We note that further works consider
time-dependent scattering \cite{snyman:08} or spin-dependent effects in
specific models \cite{fogelstrom:00,eschrig:04,kopu:04,cuevas:01,eschrig:08}.

The derivations in this manuscript are mainly based on the ideas
presented in \cite{nazarov:99} and \cite{cottet:09}. Thus the results
represent a generalizations of the Nazarov boundary condition for
spin-dependent problems.

Throughout the paper we indicate all operators with non-trivial
structure in more than one subspace by their decorations according to
table \ref{dec}. We will use $\sigma/\tau/\Sigma$ for the
Pauli-matrices in Spin-/Nambu-/Direction (of propagation)-space
respectively.  We work in the Spinor-basis
$\Psi\dg=(\psi\dg\ua,\psi\dg\da,-\psi\da,\psi\ua)$. Thus, we defined
the time-reversal operator as $R=-i\sigma_2R_0$, where for any
$\bar{A}$ the operator $R_0$ yields the complex conjugate
($R_0\bar{A}=\bar{A}^*$).

In this basis the transfer matrix $\tilde{{\cal M}}$ that transfers
states $(\varphi^{L/R})^{\rm
  T}=\begin{pmatrix}(\varphi^{L/R})^e,(\varphi^{L/R})^h\end{pmatrix}$
($L/R$ label the left/right side of the interface, $e/h$ labels the
electron/hole component) like $\varphi^R=\tilde{{\cal M}} \varphi^L$
has the following structure in Nambu space: 
\begin{align}
  \tilde{{\cal M}}=\begin{pmatrix}\bar{\bar{{\cal
          M}}}^e&0\\0&\bar{\bar{{\cal
          M}}}^h\end{pmatrix}=\begin{pmatrix}\bar{\bar{{\cal
          M}}}^e&0\\0&-i\sigma_2(\bar{\bar{{\cal
          M}}}^e)^*i\sigma_2\end{pmatrix}. 
\end{align}
The transfer matrix is equivalent to the scattering matrix and can be
found using the unitarity condition. As shown in \cite{cottet:09}, the
transfer matrix $\tilde{M}$ relating the ballistic Green's functions
like $\hat{g}^R=\tilde{M}\hat{g}^L\tilde{M}\dg$ is defined as 
\begin{align}
  \tilde{M}=\begin{pmatrix}\bar{\bar{{\cal
          M}}}^e&0\\0&\Sigma_1\bar{\bar{{\cal
          M}}}^h\Sigma_1\end{pmatrix}. 
\end{align}
Thus assuming the decomposition $\bar{\bar{{\cal M}}}^e=(\acute{{\cal
    M}}^e)'+(\acute{{\cal M}}^e)''\vec{m}\grave{\vec{\sigma}}$ for the
electronic transfer matrix, with the magnetization unit vector
$\vec{m}$, we find the compact form  
\begin{align}\label{mgeneral}
  \tilde{M}=\acute{M}'+\acute{M}''\check{\kappa}.
\end{align}
Here we defined $\acute{M}'=(\acute{{\cal M}}^e)'$,
$\acute{M}''=(\acute{{\cal M}}^e)''$ and
$\check{\kappa}=\tau_3\otimes\vec{m}\vec{\sigma}$. To find the
matrices $\acute{M}'$ and $\acute{M}''$, we start with the most
general form of the electronic transfer  matrix (see
e.g. \cite{cottet:09}): 
\begin{align}\label{transfere}
\bar{\bar{{\cal M}}}_n^e=
\begin{pmatrix}
\frac{i}{\sqrt{\grave{T}_{n}}}e^{i\grave{\varphi}_{n}/2}
&-i\sqrt{\frac{1-\grave{T}_{n}}{\grave{T}_{n}}}e^{i\grave{\chi}_{n}/2}\\
i\sqrt{\frac{1-\grave{T}_{n}}{\grave{T}_{n}}}e^{-i\grave{\chi}_{n}/2}
&\frac{-i}{\sqrt{\grave{T}_{n}}}e^{-i\grave{\varphi}_{n}/2}
\end{pmatrix}
\end{align}
The index $n$ labels the transport channel. We assumed the transfer
matrix to be already diagonalized in channel space\cite{beenakker:97},
which is always possible since the isotropic Greens functions are
structureless in that space.  Note that the transfer matrix includes
channel-dependent polarizations of the transmission coefficients $P_n$, spin-dependent phase shifts
(spin mixing angles) $\delta\phi^{L/R}_n$ and even the
spin-quantization axes $\vec m_n$. Explicitly we defined
$\grave{T}_{n}=T_n(1+P_n\vec{m}_n\vec{\sigma})$ and
$\grave{\phi}^{L/R}_{n}=\phi^{L/R}_n+(\delta\phi^{L/R}_n/2)\vec{m}_n\vec{\sigma}$,
while $\grave{\varphi}_{n}=\grave{\phi}^R_{n}+\grave{\phi}^L_{n}$ and
$\grave{\chi}_{n}=\grave{\phi}^R_{n}-\grave{\phi}^L_{n}$.

The desired form of $\bar{\bar{{\cal M}}}^e$, Eq.~\eqref{mgeneral} is
found using the matrix structure of $\vec{m}\vec{\sigma}$ to rewrite
the prefactor of the matrix elements in Eq.~\eqref{transfere} according to
\begin{align}
&\frac{1}{\sqrt{\grave{T}}}=\frac{1}{\sqrt{T'}}(p_+-p_-\vec{m}\vec{\sigma})\label{thelp}\\
&\sqrt{\frac{1-\grave{T}}{\grave{T}}}=\frac{1}{\sqrt{T'}}(t_+-t_-\vec{m}\vec{\sigma})\label{rhelp}
\end{align}
with
\begin{align}
  &T'=T(1-P^2)\\
  &p_{+/-}=\sign{P}^{0/1}\sqrt{\frac12(1\pm\sqrt{1-P^2})}\label{ps}\\
  &t_{+/-}=\sign{P}^{0/1}\sqrt{\frac12(1-T'\pm\sqrt{(1-T')^2-P^2})}.\label{ts}
\end{align}
Note that all these quantities (transmission, magnetization,
polarization, phases and phase shifts) are diagonal but non-trivial
matrices in the channel subspace. We expand Eq.~\eqref{mgeneral} with
the help of Eqs.~\eqref{transfere}, \eqref{thelp}, and \eqref{rhelp} and find 
\begin{align}
M'_{11}&=(\acute M')_{11}=\frac{i}{\sqrt{T'}}e^{i\varphi/2}(p_+c_{\varphi}-ip_-s_{\varphi}) \label{m11s}\\
M''_{11}&=(\acute M'')_{11}=\frac{i}{\sqrt{T'}}e^{i\varphi/2}(-p_-c_{\varphi}+ip_+s_{\varphi}) \label{m11ss}\\
M'_{12}&=(\acute M')_{12}=\frac{-i}{\sqrt{T'}}e^{i\chi/2}(t_+c_{\chi}-it_-s_{\chi}) \label{m12s}\\
M''_{12}&=(\acute M'')_{12}=\frac{-i}{\sqrt{T'}}e^{i\chi/2}(-t_-c_{\chi}+it_+s_{\chi}). \label{m12ss}
\end{align}
We defined $c_\alpha=\cos(\delta\alpha/4)$ and $s_\alpha=\sin(\delta\alpha/4)$.
All other components are related by further symmetries (equivalent to the unitarity of the scattering matrix):
\begin{align*}
M'_{22}=(M'_{11})^*\,\,&;\,\,M''_{22}=(M''_{11})^*\\
M'_{21}=(M'_{12})^*\,\,&;\,\,M''_{21}=(M''_{12})^*.
\end{align*}

\begin{table}
\begin{tabular*}{0.8\linewidth}{c @{\extracolsep{\fill}} | c | c | c | c | c | c | c | c }
\hline\hline
           & $\grave{A}$ & $\acute{A}$ &  $\hat{A}$  & $\check{A}$ & $\breve{A}$ &  $\bar{\bar{A}}$  & $\tilde{A}$ & $\bar{A}$  \\
\hline\hline
spin       &  \ding{51}  &  \ding{55}  &  \ding{51}  &  \ding{51}  &  \ding{51}  &     \ding{51}     &  \ding{51}  &  \ding{51} \\\hline
Nambu      &  \ding{55}  &  \ding{55}  &  \ding{51}  &  \ding{51}  &  \ding{51}  &     \ding{55}     &  \ding{51}  &  \ding{51} \\\hline
direction  &  \ding{55}  &  \ding{51}  &  \ding{55}  &  \ding{55}  &  \ding{55}  &     \ding{51}     &  \ding{51}  &  \ding{51} \\\hline
Keldysh    &  \ding{55}  &  \ding{55}  &  \ding{51}  &  \ding{55}  &  \ding{51}  &     \ding{55}     &  \ding{55}  &  \ding{51} \\\hline
channel    &  \ding{51}  &  \ding{51}  &  \ding{55}  &  \ding{51}  &  \ding{51}  &     \ding{51}     &  \ding{51}  &  \ding{51} \\
\hline\hline
\end{tabular*}
\caption{Explanation of decorations used throughout the paper. The symbols \ding{51} and \ding{55} indicate that the operator has or has no structure in the particular subspace.}
\label{dec}
\end{table}

The matrix-current in the isotropization zone on the left/right side
of the contact is found in \cite{nazarov:99,cottet:09}. For simplicity
we define $\hat{G}=\hat{G}^{L/R}$, $\hat{G}'=\hat{G}^{R/L}$, and
$\bar{K}=\bar{K}^{L/R}$, where the indices $L/R$ label the left/right
side of the interface respectively and $\hat{G}^{L/R}$ are the
quasiclassical Green's functions of either sides. With these
definitions and defining the conductance quantum $G_Q=e^2/h$, the matrix-current ($\hat{I}=\hat{I}^{L/R}$) is   
\begin{align}
  \hat{I}=4G_Q\rm{Tr}_{n,s}[(1+\hat{G}
  \bar{K})^{-1}(\hat{G}\Sigma_3\pm1)\mp\frac12],\label{matcurr} 
\end{align}
with
\begin{align}
  &\bar{K}=(\tilde{M}\dg)^{\pm1} \hat{G}'\tilde{M}^{\pm1}.& \label{krare}
\end{align}
The index $s$ labels the trace in direction subspace. For later usage
we mention that due to $\tilde{M}\dg \Sigma_3
\tilde{M}=\tilde{M}\Sigma_3 \tilde{M}\dg=\Sigma_3$ and the
normalization condition $(\hat{G})^2=1$, it is: 
\begin{align}\label{matstructur}
  (\Sigma_3 \bar{K})^2=1.
\end{align}
The inverse of the 2x2 matrix in the direction subspace in
Eq.~\eqref{matcurr} is found in terms of components $\breve{K}_{ij}$
of $\bar{K}$ as follows. We define 
\begin{flalign}
  &\bar D^{-1}=(1+\hat{G} \bar{K})^{-1}&\nonumber\\
  &=\begin{pmatrix}
    1+\hat G \breve K_{11}& \hat G \breve K_{12}\\
    \hat G \breve K_{21}&1+\hat G \breve K_{22}
  \end{pmatrix}^{-1}\hspace{-10pt}=\hspace{-2pt}
  \begin{pmatrix}
    (\breve D^{-1})_{11}&(\breve D^{-1})_{12}\\
    (\breve D^{-1})_{21}&(\breve D^{-1})_{22}
  \end{pmatrix}.&\raisetag{1.5\baselineskip}
\end{flalign} 
Performing the trace over the $s$-subspace in Eq.~\eqref{matcurr} we find:
\begin{flalign}\label{I2} 
&\hat{I}=4G_Q\rm{Tr}_{n}[(\breve D^{-1})_{11}(\hat G\pm1)-(\breve D^{-1})_{22}(\hat G\mp1)\mp 1].&\raisetag{0.8\baselineskip}
\end{flalign}
The entries of the 2x2 matrix $\bar{D}^{-1}$ are found by direct inversion:
\begin{flalign}
&(\breve D^{-1})_{11}\hspace{-2pt}=\hspace{-2pt}(1+\hat{G}\breve{K}_{11}-\hat{G}\breve{K}_{12}(1+\hat{G}\breve{K}_{22})^{-1}\hat{G}\breve{K}_{21})^{-1}&\raisetag{0.8\baselineskip}\label{d11}\\
&(\breve D^{-1})_{22}\hspace{-2pt}=\hspace{-2pt}(1+\hat{G}\breve{K}_{22}-\hat{G}\breve{K}_{21}(1+\hat{G}\breve{K}_{11})^{-1}\hat{G}\breve{K}_{12})^{-1}\hspace{-2pt}.&\raisetag{0.8\baselineskip}\label{d22}
\end{flalign}
Using the matrix structure of $\bar{K}$ (Eq.~\eqref{matstructur}), i.e. $\breve K_{21}^{-1}\breve{K}_{22}=\breve{K}_{11}\breve{K}_{21}^{-1}$ and $\breve{K}_{21}^{-1}+\breve{K}_{12}=\breve{K}_{11}\breve{K}_{21}^{-1}\breve{K}_{22}$, Eqs.~\eqref{d11} and \eqref{d22} can be decomposed according to:
\begin{align}
&(\breve D^{-1})_{11}=\breve{C}\breve{A}^{-1}\label{d11n}\\
&(\breve D^{-1})_{22}=\breve A^{-1}( \breve K_{21}^{-1}\hat G+\hat G \breve K_{11} \breve K_{21}^{-1}\hat G)\label{d22n}
\end{align}
with
\begin{align}
&\breve C=(\hat G \breve K_{21})^{-1}(1+\hat G \breve K_{22})= \breve K_{21}^{-1}\hat G+ \breve K_{11} \breve K_{21}^{-1}\label{dc}\\
&\breve A=\breve K_{21}^{-1}\hat G+ \hat G  \breve K_{11} \breve K_{21}^{-1}\hat G+ \breve K_{21}^{-1} \breve K_{22}\nonumber\\
&\phantom{\breve A =}+\hat G \breve K_{11} \breve K_{21}^{-1} \breve K_{22}-\hat G \breve K_{12})^{-1}\nonumber\\
&\phantom{\breve A }=\breve C+\hat G\breve C\hat G.\label{da}
\end{align}
Plugging Eqs.~\eqref{d11n}, \eqref{d22n}, \eqref{dc}, and \eqref{da} into
Eq.~\eqref{I2}, replacing the last summand $\mp 1$ in Eq.~\eqref{I2}
by $A^{-1}A$ and using the relation $[A^{-1},\hat G]=0$ it is straight
forward to show that \cite{eqn}
\begin{align}\label{bc}
  &\hat I=4G_Q\text{Tr}_n[(1\pm \breve{K}_{11})(\breve{K}_{21})^{-1},\nonumber\\*
  &\hphantom{{}\hat I=4G_Q\sum_n[}\{(\hat
    G+\breve{K}_{11})(\breve{K}_{21})^{-1},\hat G\}^{-1}(\hat
    G\pm1)]. 
\end{align}
This is the most general form of the desired boundary condition, and
thus the main result of the paper. To solve physical problems, we
still need to find the components of $\bar K$ to be plugged into
Eq.~\eqref{bc}. For this we plug Eq.~\eqref{mgeneral} into Eq.~\eqref{krare}
(note that $\det{\tilde M}=1$): 
\begin{flalign}\label{kexpand}
&\bar{K}\hspace{-1.5pt}=\hspace{-1.5pt}((\acute{M}')\dg)^{\pm 1} (\acute{M}')^{\pm 1}\hat{G}'\hspace{-1.5pt}+\hspace{-1.5pt}((\acute{M}'')\dg)^{\pm 1} (\acute{M}'')^{\pm1}\check \kappa \hat{G}'\check \kappa&\nonumber\\*
&\hphantom{\bar{K}\hspace{-1.5pt}}+\hspace{-1.5pt}((\acute{M}')\dg)^{\pm 1} (\acute{M}'')^{\pm 1}\hat{G}'\check \kappa\hspace{-1.5pt}+\hspace{-1.5pt}((\acute{M}'')\dg)^{\pm1} (\acute{M}')^{\pm1}\check \kappa \hat{G}'.&\raisetag{0.8\baselineskip}
\end{flalign}
The last two summands are related by the symmetry
$(((\acute M')\dg)^{\pm 1} (\acute M'')^{\pm 1})\dg=((\acute
M'')\dg)^{\pm 1} (\acute M')^{\pm 1}$.
Plugging Eqs.~\eqref{m11s}, \eqref{m11ss}, \eqref{m12s} and
\eqref{m12ss} into Eq.~\ref{kexpand} using all the symmetries one
finds the following for the desired components of $\bar{K}$:
\begin{align}
  &\breve K_{11}=K'_{11}\hat G'+K''_{11}\check \kappa G'\check \kappa+K'''_{11}\hat G'\check \kappa+K'''^*_{11}\check \kappa \hat G'\\
  &\breve K_{21}=K'_{21}\hat G'+K''_{21}\check \kappa \hat G'\check \kappa+K'''_{21}\{\hat G',\check \kappa\}
\end{align}
with
\begin{widetext}
\begin{align}
    &(K^{'/''}_{11})^L=-\frac12+\frac{1}{T'} \left[ 1\pm\frac12\left\{\sqrt{1-P^2}\cos(\delta\varphi/2)+\sqrt{(1-T')^2-P^2}\cos(\delta\chi/2)\right\}\right]\\
                 &(K'''_{11})^L=\frac{1}{T'} \left[ -P+\frac{i}{2}\left\{\sqrt{1-P^2}\sin(\delta\varphi/2)-\sqrt{(1-T')^2-P^2}\sin(\delta\chi/2) \right\}\right]\\
                 &(K^{'/''}_{21})^L=-\frac{1}{T'}e^{i\phi^L}\left[(p_+t_++p_-t_-)\cos(\delta\phi^L/2)\pm(p_+t_+-p_-t_-)\cos(\delta\phi^R/2)\right.\nonumber\\*
                 &\phantom{{}(K^{'/''}_{21})^L=-\frac{1}{T'}e^{i\phi^{L}}(} 
                   \left.-i\left\{(p_-t_++p_+t_-)\sin(\delta\phi^L/2)\mp(p_+t_--p_-t_+)\sin(\delta\phi^R/2) \right\}\right]\\
                 &(K'''_{21})^L=\frac{1}{T'}e^{i\phi^L}\left[(p_+t_-+p_-t_+)\cos(\delta\phi^L/2)-i(p_+t_++p_-t_-)\sin(\delta\phi^L/2)\right]
\end{align}
Here we partly used Eqs.~\eqref{ps} and \eqref{ts}. Due to the structure of
the transfer-matrix the right-side coefficients are given by
$(K^{'/''/'''}_{ab})^R=R_0(K^{'/''/'''}_{ab})^L|_{L\leftrightarrow  R}$. 
Now we can express the full boundary condition with
($\phi=\phi^{L/R},\,\phi'=\phi^{R/L},\,\delta\phi=\delta\phi^{L/R},\,\delta\phi'=\phi^{R/L}$)
\begin{align}
  &\breve K_{11}=(\hat G'+\check \kappa \hat G'\check \kappa)\left[\frac{1-P\check \kappa}{T'}-\frac12\right]\nonumber\\*
  &\hphantom{{}\breve K_{11}}+\frac{1}{2T'}(\hat G'-\check \kappa \hat
    G'\check \kappa) \left[\sqrt{1-P^2}\,e^{\pm i\check \kappa(\delta\phi+\delta\phi')/2}+\sqrt{(1-T')^2-P^2}\,e^{\pm i\check \kappa(\delta\phi-\delta\phi')/2}\right]\label{bc1}\\
  &\breve K_{21}=-\frac{1}{T'}e^{\pm i\phi}\left\{(\hat G'+\check \kappa
    \hat G'\check \kappa)\left[p_+t_++p_-t_--\check
    \kappa(p_+t_-+p_-t_+)\right]\,e^{\pm i\check \kappa\,\delta\phi/2}\right.\nonumber\\*
  &\left.\hphantom{{}\breve K_{21}=-\frac{1}{T'}e^{\pm i\phi^{L/R}}}
    +(\hat G'-\check \kappa \hat G'\check \kappa) 
    \left[(p_+t_+-p_-t_-)\cos(\delta\phi'/2)\pm i(p_+t_--p_-t_+)\sin(\delta\phi'/2)\right]\right\}\label{bc2}.
\end{align}
\end{widetext}
This completes the central result of our work. Plugging
Eqs.~\eqref{bc1} and \eqref{bc2} into Eq.~\eqref{bc} yields the dirty
limit matrix current through a contact where both sides ($\hat G^R$ and
$\hat G^L$) can have a non-trivial structure in Nambu and spin space
and the spin-dependent interface effects i.~e. spin mixing and spin
filtering are treated exactly. It completes the discrete version of
the Usadel equation \cite{nazarov:99} to yield a complete scheme
to describe a large variety of superconductor-ferromagnet
heterostructures, with external fields under equilibrium and
non-equilibrium conditions.

In the remainder of the paper we discuss several practically relevant
simplifications of the general boundary conditions. At first we note
that the Nazarov boundary condition \cite{nazarov:99} is reproduced for
$\delta\phi=\delta\phi'=P=0$, where $\breve K_{11}=(2-T)\hat G'/T$ and
$\breve K_{21}=-2\sqrt{1-T}\hat G'/T$. Thus, with
$[\hat G',(\frac{T}{2(T-2)}\{\hat G',\hat G\}-1)^{-1}]=0$,
Eq.~\eqref{bc} reduces to the well-known result
\begin{align}
\hat I=4\,G_Q\sum_n\frac{T_n[\hat G',\hat G]}{4+T_n(\{\hat G',\hat G\}-2)}.
\end{align}
In many cases tunnel barriers (with or without a magnetic structure)
are an experimentally realized and  we
will at first take the tunnel limit ($T_n\ll1$) by linear
expansion of Eqs.~\eqref{bc1} and \eqref{bc2}: 
\begin{align}\label{tunnel}
&T'\breve K_{11/21}\approx \breve K_{11/21}^0+\breve K_{11/21}^1T
\end{align}
with
\begin{widetext}
\begin{align}
&\breve K_{11}^0=(1-P\check \kappa)(\hat G'+\check \kappa \hat G'\check \kappa)+(\hat G'-\check \kappa \hat G'\check \kappa)\sqrt{1-P^2}\cos{(\delta\phi'/2)}\,e^{\pm i\check \kappa\,\delta\phi/2}\\
&\breve K_{11}^1 = 
  -\frac12\left[(1-P^2)(\hat G'+\check \kappa \hat G'\check \kappa)+\sqrt{1-P^2}(\hat G'-\check \kappa \hat G'\check \kappa)\,e^{\pm i\check \kappa(\delta\phi-\delta\phi')/2} \right]\\
&\breve K_{21}^0=-e^{\pm i\phi}\left[(1-P\check \kappa)(\hat G'+\check \kappa \hat G'\check \kappa)e^{\pm i\check \kappa\,\delta\phi/2}+\sqrt{1-P^2}(\hat G'-\check \kappa \hat G'\check \kappa)\cos{(\delta\phi'/2)}\right]\label{tunnel3}\\
&\breve K_{21}^1=\frac{e^{\pm i\phi}}{2}\left[(1-P^2)(\hat G'+\check \kappa \hat G'\check \kappa)e^{\pm i\check \kappa\,\delta\phi/2}+\sqrt{1-P^2}(\hat G'-\check \kappa \hat G'\check \kappa)(\cos{(\delta\phi'/2)}\mp iP\sin({\delta\phi'/2)})\right].
\end{align} 
Plugging Eq.~\eqref{tunnel} into Eq.~\eqref{bc} results in
\begin{align}
\hat I=4G_Q\text{Tr}_n& [ \breve K_{11}^0(\breve K_{21}^0)^{-1},\{\breve K_{11}^0(\breve K_{21}^0)^{-1},\hat G \}^{-1} (1-\{(\hat G(1-P^2)+\breve K_{11}^1-\breve K_{11}^0(\breve K_{21}^0)^{-1}\breve K_{21}^1)(\breve K_{21}^0)^{-1},\hat G \}\times\nonumber\\*
\times &\{\breve K_{11}^0(\breve K_{21}^0)^{-1},\hat G \}^{-1}T)(1\pm \hat G) ] + [ (\pm(1-P^2)+\breve K_{11}^1-\breve K_{11}^0(\breve K_{21}^0)^{-1}\breve K_{21}^1)(\breve K_{21}^0)^{-1},\nonumber\\*
&\{\breve K_{11}^0(\breve K_{21}^0)^{-1},\hat G \}^{-1}(1\pm \hat G) ] T.
\end{align} 
\end{widetext}
Using the relation $\forall\,\alpha\in\mathbb{C}:\,e^{-i\check
  \kappa\alpha}[\hat G',\check \kappa]=[\hat G',\check
\kappa]e^{i\check \kappa\alpha}$ one finds
\begin{flalign}
&\breve K_{11}^0(\breve K_{21}^0)^{-1}=-e^{\mp i(\phi+\check \kappa\,\delta\phi/2)}&\\
&\breve K_{11}^1+e^{\mp i\check \kappa\,\delta\phi/2}\breve K_{21}^1=&\nonumber\\*
&=\frac{\pm i}{2}\sqrt{1-P^2}[\hat G',\check \kappa](1-P\check \kappa)\sin{(\delta\phi'/2)}=\pm \breve K^1\label{k1},&\raisetag{1.1\baselineskip}
\end{flalign}
and thus
\begin{flalign}\label{bct}
\hat I\hspace{-0.5pt}&=\hspace{-0.5pt}4G_Q\sum_n[e^{\mp i\check \kappa_n\,\delta\phi_n/2},\{e^{\mp i\check \kappa_n\,\delta\phi_n/2},\hat G\}^{-1}(\hat G\pm 1)]&\nonumber\\*
+&\hspace{-0.5pt}T_n([e^{\mp i\check \kappa_n\,\delta\phi_n/2},\{e^{\mp i\check \kappa_n\,\delta\phi_n/2},\hat G\}^{-1}\{ (\hat G(1-P_n^2)\pm \breve K^1_n)& \nonumber\\*
\times&\hspace{-0.5pt}(\breve K_{21}^0)^{-1}_n,\hat G\}\{e^{\mp i\check \kappa_n\,\delta\phi_n/2},\hat G\}^{-1}(\hat G\pm 1)]&\nonumber\\*
-&\hspace{-0.5pt}[(1\hspace{-0.5pt}-\hspace{-0.5pt}P_n^2\hspace{-0.5pt}+\hspace{-0.5pt}\breve K^1_n)(\breve K_{21}^0)^{-1}_n\hspace{-1.5pt},\hspace{-1.5pt}\{e^{\mp i\check \kappa_n\,\delta\phi_n/2}\hspace{-1.5pt},\hat G\}^{-1}(\hat G\pm 1)]).&\raisetag{0.8\baselineskip}
\end{flalign}
Eq.~\eqref{bct} together with Eqs.~\eqref{k1} and \eqref{tunnel3}
represents the tunnel limit of the full boundary condition
(eq. (\ref{bc})). In the case of a ferromagnetic insulator $T_n\equiv0$
only the first line in the Eq.~\eqref{bct} remains. Further simplifications are found assuming small spin-mixing. We note that
\begin{align}\label{inv}
&\left(\{\hat G',\check \kappa\}(\check \kappa-P)+[\hat G',\check \kappa]\check \kappa\sqrt{1-P^2}\right)^{-1}=\nonumber\\*
=&\frac{\{\hat G',\check \kappa\}(\check \kappa+P)+[\hat G',\check \kappa]\check \kappa\sqrt{1-P^2}}{4(1-P^2)}.
\end{align} 
By expansion of Eq.~\eqref{bct} to first order in the spin-dependent phase shifts ($\delta\phi_n,\,\delta\phi'_n$), using Eq.~\eqref{inv} we find:
\begin{align}
2\hat I=G_Q\sum_n&\left[\pm T_n\left((1+\sqrt{1-P_n^2})\hat G'+P_n\{\hat G',\check \kappa_n\} \right.\right.\nonumber\\*
+&\left.\left. (1-\sqrt{1-P_n^2})\check \kappa_n \hat G'\check \kappa_n\right)-2i\,\delta\phi_n ,\hat G\right].\label{bcold}
\end{align}
This is the boundary condition as it is used in
\cite{machon:13,machon:14} to calculate thermoelectric effects in
ferromagnet-superconductor heterostructures. 

In conclusion we have found a new boundary condition for isotropic
quasicalcassical Greens functions that generalizes the Nazarov
boundary condition in the quantum circuit theory to spin-dependent
problems. It is presented in Eq.~\eqref{bc} supplemented by
Eqs.~\eqref{bc1} and \eqref{bc2}. The new boundary condition captures
e.g. non-collinear ferromagnetes and triplet superconductors as
terminals and fully accounts for arbitrary transmissions,
polarizations and spin-mixing effects of the scattering region. In the
tunnel limit the boundary condition can be simplified to
Eq.~\eqref{bct}, still fully accounting for arbitrary
spin-mixing. This case also comprises ferromagnetic insulators by
setting $T_n=0$ given by the first line of Eq.~\eqref{bct}. The next
drastic yet non-trivial simplification of the boundary condition is to
linearize in the spin-dependent phase shifts given in
Eq.~\eqref{bcold}. In this case the boundary condition takes a form of
a commutator (similar to \cite{huertas:02} and in \cite{nazarov:99})
and the transport coefficients just enter in averaged interface
parameters $T_n$, $P_n$ and $\delta\phi^{L/R}_n$. Our results pave the
way to investigate spin-dependent charge and heat transport and spin
transport in superconducting heterostructures as well as
unconventional correlations like odd-frequency triplet pairing using
the most general boundary condition based on a fully microscopic
derivation. Another application is to verify the details of
microscopic models describing the intrinsic magnetic structure and
magnetization behavior of e.g ferromagnetic insulators.

We thank M. Eschrig and W. Timmermann for very useful discussions.
We acknowledge financial support from the DFG through BE 3803/3 and
SPP 1538 and the Baden-W\"urttemberg Foundation within
the Research Network of Excellence "Functional Nanostructures''.

\end{document}